\documentclass[aps,pra,superscriptaddress]{revtex4}
\usepackage[dvips]{graphicx}
\usepackage{amssymb}
\usepackage{amsmath}
\usepackage{hyperref}

\begin{document}

%
\title{Chaotic laser based physical random bit streaming system with a
  computer application interface}

\newcommand{\micron}{\mu\mbox{m}}

\newcommand{\NTTCS}{\affiliation{NTT Communication Science
    Laboratories, NTT Corporation, 2-4 Hikaridai, Seika-cho,
    Soraku-gun, Kyoto 619-0237, Japan}}
\newcommand{\TELECOGNIX}{\affiliation{Telecognix Corporation, 58-13
    Shimooji-cho, Yoshida, Sakyo-ku, Kyoto, 606-8314 Japan}}
\newcommand{\KANAZAWA}{\affiliation{Faculty of Mechanical Engineering,
    Institute of Science and Engineering, Kanazawa University,
    Kakuma-machi, Kanazawa, Ishikawa 920-1192, Japan}}
\newcommand{\WASEDA}{\affiliation{Department of Applied Physics,
    School of Advanced Science and Engineering, Waseda University,
    3-4-1 Okubo, Shinjuku-ku, Tokyo 169-8555, Japan}}
\author{Susumu Shinohara}\email{01100101@runbox.com}\NTTCS
\author{Kenichi Arai}\NTTCS
\author{Peter Davis}\TELECOGNIX
\author{Satoshi Sunada}\KANAZAWA
\author{Takahisa Harayama}\WASEDA
\date{\today}
\begin{abstract}
We demonstrate a random bit streaming system that uses a chaotic
laser as its physical entropy source.
By performing real-time bit manipulation for bias reduction, we were
able to provide the memory of a personal computer with a constant
supply of ready-to-use physical random bits at a throughput of up to 4
Gbps.
We pay special attention to the end-to-end entropy source model
describing how the entropy from physical sources is converted into bit
entropy.
We confirmed the statistical quality of the generated random bits by
revealing the pass rate of the NIST SP800-22 test suite to be 65 \% to
75 \%, which is commonly considered acceptable for a reliable random
bit generator.
We also confirmed the stable operation of our random bit steaming
system with long-term bias monitoring.
\vskip 5mm
\noindent
\href{https://doi.org/10.1364/OE.25.006461}{doi:10.1364/OE.25.006461}
\vskip 5mm
\noindent
\copyright~2017~Optical Society of America. One print or electronic copy may be made for personal use only. Systematic reproduction and distribution, duplication of any material in this paper for a fee or for commercial purposes, or modifications of the content of this paper are prohibited.
\end{abstract}
%
%
\maketitle
%
\section{Introduction}
Since the first demonstration of giga-bit-per-second (Gbps) physical
random bit generation (RBG) using semiconductor lasers by Uchida et
al. in 2008 \cite{Uchida08}, chaotic lasers have attracted renewed
interest as an entropy source for physical RBG.
A semiconductor laser with delayed optical feedback can generate
large-amplitude, chaotic intensity fluctuations in the GHz regime
\cite{Ohtsubo}.
A chaotic laser can be viewed as an ``amplifier'' of microscopic
noises \cite{Harayama12,Sunada12}, which uses nonlinear dynamical
instability as the amplification mechanism.
Although dynamical instability itself is deterministic, the
non-deterministic randomness involved in microscopic noises results in
unpredictable output behavior after the Lyapunov time.
The practical unpredictability of a chaotic laser is demonstrated
experimentally in \cite{Sunada12}.

In the field of information and communication technology (ICT), there
is a great demand for physical (truly) random bit generators, because
true randomness plays a crucial role as regards achieving secure
communication and storage systems.
A high generation rate for physical random bits is useful for
applications based on information-theoretic security \cite{Shannon49}
such as secret sharing \cite{Shamir79,Blakley79} and secure
multi-party computation \cite{Yao82,Goldreich87}.
These applications consume very many random numbers because their
perfect security (in terms of information theory) is guaranteed by as
much true randomness (or full entropy) as original information that
one wants to encrypt.
RBG using a chaotic laser has the potential to meet the need for
high-generation-rate physical random bit generators.

With RBG using a chaotic laser, an AD converter (ADC) that resolves
the state of the laser is as important as the entropy source.
For example, digitization resolution of an ADC is directly related to
the RBG rate.
Many studies have demonstrated the enhancement of the RBG rate by
multi-bit sampling \cite{Reidler09, Hirano09, Hirano10, Kanter10,
  Argyris10, Akizawa12, Oliver13, Takahashi14, Sakuraba15}, although
in these cases bit manipulation is needed to reduce the biases of the
generated bits.
Thus far, such bit manipulation has only been performed offline.
Moreover, real-time AD conversion has only been examined in a few
studies \cite{Uchida08, Honjo09, Zhang12, Wang13}, and all of these
studies used one-bit ADCs.

The high generation rate proves its worth when it is used in a
real-time physical random bit streaming (RBS) system, which can both
perform bit manipulation for bias reduction in real time and
continuously supply random bits to the memory of a server or a
personal computer (PC).
As a step towards application in the ICT field, it is important to
examine the feasibility of such a real-time streaming system by using
currently available electronic devices.
We addressed this new challenge by integrating a chaotic laser with a
commercially available ADC board that had a data interface with a PC,
paying special attention to the end-to-end entropy source model, that
is, how the entropy from physical sources is converted into bit
entropy.
We developed software for real-time bit manipulation for bias
reduction and streaming to a PC user's memory space, and demonstrated
a physical RBS rate of up to 4 Gbps.
We emphasize that this rate was measured in terms of the throughput of
physical random bits supplied to a PC user's memory space, in contrast
to previous theoretical estimates based on offline post processing.

Our RBS system allows us to easily store and access physical random
bits in a PC.
We used this advantage to carry out a comprehensive assessment of
physical random bits.
In particular, we evaluated the pass rate of the NIST test suite for
random number generators \cite{SP800-22}.
For various parameter values of the bit manipulation algorithm, we
found that the pass rates were from 65 \% to 75 \%, which are
comparable to the pass rates previously evaluated for commonly-used
reliable pseudorandom bit generators \cite{Okutomi10, Yamaguchi10,
  Lihua15}.

\section{Random bit streaming system}
\subsection{Chaotic laser chip}
\begin{figure}[b]
\centering
\includegraphics[width=9cm]{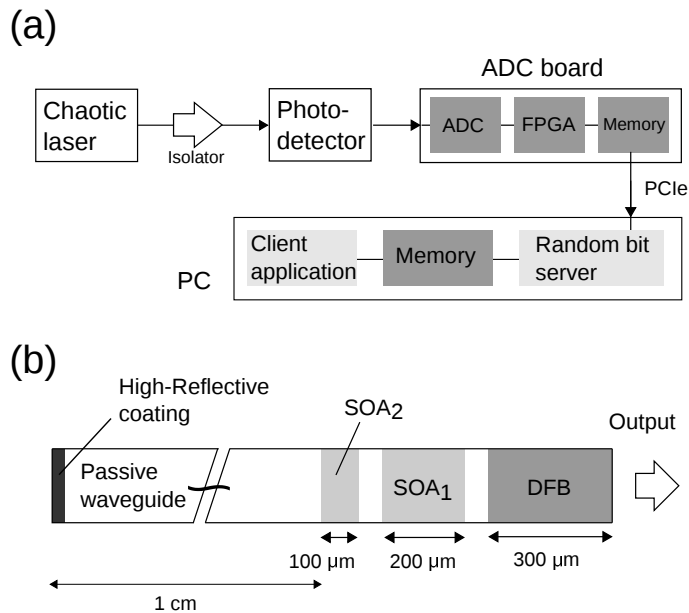}
\caption{(a) Schematic diagram of random bit streaming system. (b)
  Structure of chaotic laser chip.
}
\label{fig:setup}
\end{figure}

A schematic diagram of our RBS system is shown in
Fig. \ref{fig:setup}(a).
The physical entropy source is a monolithically integrated chaotic
laser chip, whose structure is sketched in Fig. \ref{fig:setup}(b).
It consists of a single frequency distributed feedback (DFB) laser
emitting at a wavelength of 1550 nm, two semiconductor optical
amplifiers (SOA$_1$ and SOA$_2$), and a passive waveguide one of whose
edges is coated with high reflectivity film.
The passive waveguide works as an external cavity for delayed optical
feedback to the DFB laser, and the two SOAs are used to tune the
strength and phase of the feedback.
Our chaotic laser chip is similar to those studied in \cite{Sunada12,
  Harayama11, Takahashi14}, except that our chip has no integrated
photodiode.
The light output intensity of the chip exhibits large-amplitude,
chaotic oscillations when the external cavity is sufficiently long and
the feedback is sufficiently strong.
We set the length of the external cavity at 1.03 cm to ensure that
there were broadband chaotic oscillations in the GHz regime
\cite{Takahashi14}.

The injection currents for the DFB laser (denoted by $I_{DFB}$) and
two SOAs (denoted by $I_{SOA1}$ and $I_{SOA2}$), and the temperature
of the chip were adjusted with a current-temperature controller (ILX
Lightwave LDC-3900).
For the experiments reported in this paper, we fixed the injection
currents at $I_{DFB}$ $=$ 20.02 $\pm$ 0.01 mA (the lasing threshold
was around 12 mA), $I_{SOA1}$ $=$ 3.08 $\pm$ 0.01 mA, and $I_{SOA2}$
$=$ 2.45 $\pm$ 0.01 mA.
The temperature was fixed at 20.0 $\pm$ 0.1 $^{\circ}$C.
The output from the chaotic laser chip was converted to an electronic
signal with a high-speed AC-coupled photodetector (New Focus, 1544-B, 12
GHz bandwidth).
We used an optical fiber isolator to avoid unwanted feedback to the
laser as shown in Fig. \ref{fig:setup}(a).

\subsection{AD conversion and data streaming}
\begin{figure}[!t]
\centering
\includegraphics[width=15cm]{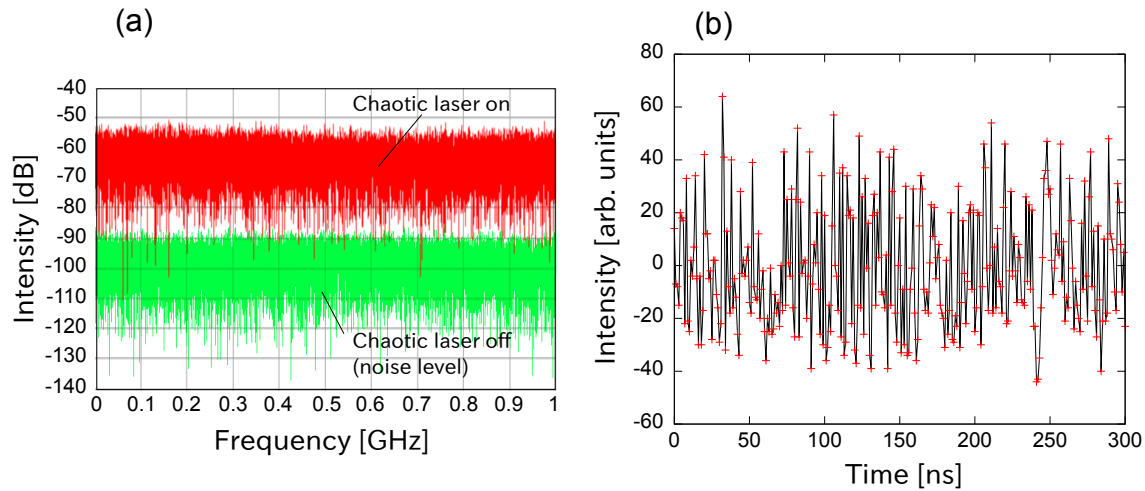}
\caption{(a) Radio-frequency spectrum and (b) time series of the
  chaotic laser output measured using the ADC board.
In (a), the noise level is given for comparison.
In (b), the sampling rate is set at 1 GSps.
}
\label{fig:signal}
\end{figure}
The electronic signal from the photodetector was transferred to an ADC
board (SP-Devices, ADQ-412-4G, 2-GHz analog bandwidth, 12-bit
digitization resolution) through a coaxial cable.
The ADC board can sample signals at a rate of up to 2
giga-samples-per-second (GSps), and sampled data are transferred to a
PC via a PCI express (Gen2$\times$8) interface.
Figure \ref{fig:signal}(a) shows radio-frequency spectra with the
chaotic laser turned on and turned off (i.e., noise level).
The spectral data were acquired with software called ADCaptureLab
(SP-Devices) in the 2-GHz bandwidth mode and with full 12-bit
resolution.
In Fig. \ref{fig:signal}(a), we can see that the intensity increased
by 35 dB compared with the noise level, and it exhibits a flat
dependence on frequency as with white noise.
We will show that these characteristics are suitable for the physical
entropy source of RBG.

The 12-bit range of the ADC corresponds to a fixed peak-to-peak
voltage of 800 mV.
We customized the Field-Programmable-Gate-Array (FPGA) on the ADC
board so that only 8-bit data were transferred to the PC, where any
eight contiguous bits can be selected by setting an FPGA parameter
from the PC, and this defines the signal range of the ADC.
For the experimental data reported in this paper, we arranged it so
that the eight bits from the lowest 4th to 11th bits were transferred
to the PC (namely, we discarded the most significant bit and the
lowest three bits of the full twelve bits), so that the signal range
of the ADC board fitted the dynamic range of the input signal.
Moreover, by programming the FPGA, we set the sampling rate of the ADC
at 1 GSps (i.e., 1-ns sampling interval).
This is because the timescale needed for amplifying microscopic noises
to a macro-scale has been measured to be around 1 ns for a similar
chaotic laser chip \cite{Sunada12}.
That is, the statistical distribution of intensity values after 1 ns
converges close to the asymptotic distribution.
In this way, the unpredictability of a sampled data sequence is
guaranteed by the physical mechanism of the microscopic noise
amplification by nonlinear dynamical instability \cite{Sunada12}.
With the above settings for the FPGA, the 8-bit data were transferred
to the PC at 1 GSps, yielding a data transfer rate of 8 Gbps.
The data bandwidth of the PCI express Gen2 $\times$8 ($\lessapprox$
25.6 Gbps) is sufficient for this data transfer.

\subsection{Real-time bit manipulation at PC}

\begin{figure}[!t]
\centering
\includegraphics[width=7cm]{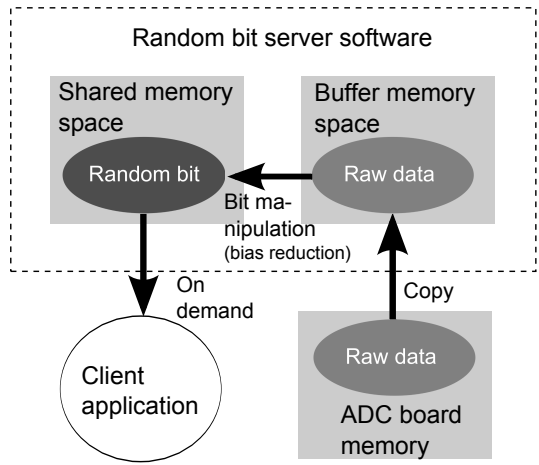}
\caption{
Diagram of data flow managed with random bit server software.
}
\label{fig:data_flow}
\end{figure}

We developed random bit server software for transferring sampled raw
data to the memory of the PC, and converting them to random bits with
reduced biases in real time.
We developed the software on a PC (2 CPUs, 8-core, 2.6 GHz, 20M cache)
with a Linux OS (Ubuntu 14.04), and for data acquisition, we used API
commands for the ADC board (SP-Devices, ADQAPI).
Figure \ref{fig:data_flow} shows a diagram of the data flow managed by
random bit server software.
The ADC board stores sampled raw data in the local memory, and those
data are continuously copied to the buffer memory space of the PC at a
data streaming rate of 8 Gbps (note that the sampling rate and
digitization resolution of the ADC were fixed at 1 GSps and 8 bits,
respectively).
Figure \ref{fig:signal}(b) shows a time-series plot of sampled raw
data for the chaotic laser's output (acquired at 1 GSps).
Then, the software converts the raw data to random bits with the bit
manipulation algorithm described in Sect. \ref{sect:bit_manipulation},
and stores the generated random bits in the shared memory space.
These random bits are ready to use, in the sense that their bias is
well reduced by the bit manipulation algorithm.
The random bits are continuously supplied to the shared memory space.
As shown in Sect. \ref{sect:bit_manipulation}, we confirmed an average
streaming throughput of up to 4 Gbps.
These constantly supplied random bits can be delivered to a client
program on demand.

\section{Physical entropy sources: chaotic laser vs. electronic noise}
\label{sect:entropy_sources}
\begin{figure}[!b]
\centering \includegraphics[width=15cm]{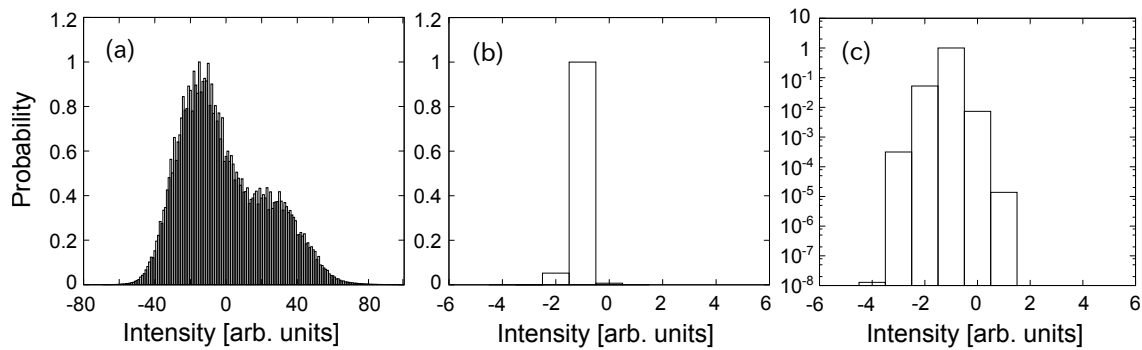}
\caption{
Typical finite-time probability distributions.
(a) Normalized probability distribution for the intensity of the
  chaotic laser.
(b) Normalized probability distribution for the intensity of the
  electronic noise.
(c) Same as (b) but the vertical axis is in log scale to enhance tiny
  probabilities.
When calculating these probability distributions, the intensity was
sampled at 1 GSps as shown in Fig. \ref{fig:signal}(b), and $1024^3$
samples were used.
}
\label{fig:histogram}
\end{figure}
When constructing a physical RBS system, it is important to clarify
the origin of the randomness, or entropy.
This point is emphasized in NIST SP800-90B \cite{SP800-90B}, for
example, which requires {\em a description of how the noise source
  works and rationale about why the noise source provides acceptable
  entropy output}.
The randomness and unpredictability of the chaotic laser outputs have
been extensively studied through physical modeling and related
numerical and real experiments \cite{Harayama11, Harayama12, Sunada12,
  Inubushi15}.
On the other hand, electronic noise is always unavoidable when a
chaotic laser is integrated with electronic components.
Thus, we first report the extent to which the electronic noise
contributes as an additional physical entropy source.

Figures \ref{fig:histogram}(a) and \ref{fig:histogram}(b) respectively
show typical finite-time probability distributions for the intensity
of the chaotic laser and for the electronic noise (for the latter, the
chaotic laser is turned off).
From the probability distribution in Fig. \ref{fig:histogram}(b)
together with direct observation of the corresponding time series, we
found that the intensity largely remained at a value of $-1$, but
sporadic fluctuation occured (in a truly noise-free case, we would
expect the intensity to remain only in a single quantization bin).
We consider this fluctuation to be mainly caused by electronic noise
intrinsic to the photodetector and the ADC board.
Figure \ref{fig:histogram}(b) shows that the intensity probability
distribution mainly spread over two bins (i.e., $-1$ and $-2$).
In Fig. \ref{fig:histogram}(c), we show a semi-log plot of the
probability distribution, where we find that the fluctuation has
distribution tails that spread over six quantization bins (this
variation corresponds to $\ln(6)/\ln(2)$ $=$ 2.58 bits).
This means that the number of bits that are not affected by the
electronic noise is estimated to be 5.42.
We need to keep this in mind when we use a chaotic laser's signals for
RBG.
For applications that do not allow use of an unknown physical entropy
source, we need to exclude those bits that are suspected of being
affected by electronic noise.

In previous studies, less attention has been paid to the treatment of
the electronic noise, and they might have constituted an additional
physical entropy source.
As far as passing the statistical tests for randomness is concerned,
no significant anomalies have been reported when some lowest bits are
included in RBG.
In Sect. \ref{sect:NIST}, we assess the quality of the generated
random bits when the electronic noise is included and when it is
excluded, for comparison.

\section{Bit manipulation for bias reduction}
\label{sect:bit_manipulation}
\begin{figure}[!b]
\centering \includegraphics[width=8cm]{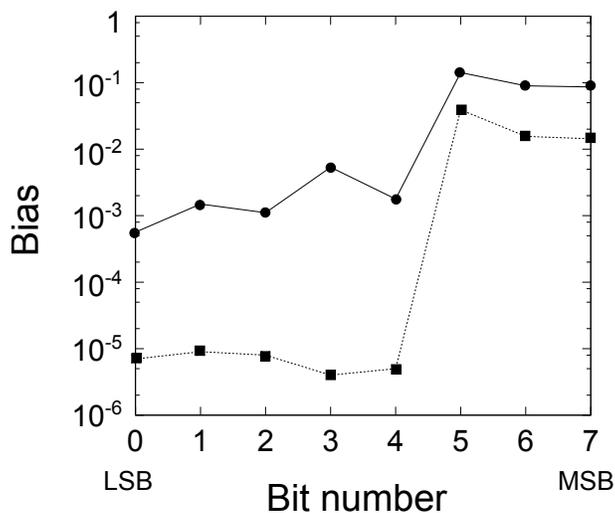}
\caption{The bias of each bit for the chaotic laser's signal, where
  1-Gbit samples were used for calculating the bias (see
  Eq. (\ref{eq:bias}) for the definition of the $j$-th bit's bias).
The filled circles ($\bullet$) are for the raw binary data
$\{A_j\}_{j=0}^7$, while the filled squares ($\blacksquare$) are for
the binary data, $\{C_0^{(s)}\}_{s=0}^7$, whose biases are reduced by
bit manipulation (see text for details).
}
\label{fig:bias}
\end{figure}

Before discussing the biases of generated bits and bit manipulation
for bias reduction, we first fix our notations.
We denote the decimal expression of the 8-bit sampled data at time $t$
$=$ $n T$ ($n$ $\in$ $\mathbf{Z}$) by $X(nT)$ $\in$ $[-128,127]$,
where $X(nT)$ takes an integer value, and the sampling interval $T$ is
fixed at $T$ $=$ 1 ns.
We denote the binary expression of $X(n T)$ by
$\{A_j^{(n)}\}_{j=0}^{7}$ with $A_j^{(n)}$ $\in$ $\{0,1\}$
$(j=0,\cdots,7)$, that is,
$X(n T)=\sum_{j=0}^7 A_j^{(n)}2^j-128$,
where $j$ $=0$ and $j$ $=$ 7 correspond to the lowest significant bit
(LSB) and the most significant bit (MSB), respectively.
We assume that for a given $j$ ($=$ 0,$\cdots$,7), $A_j^{(n)}$ is an
independent identically distributed random variable (as a function of
$n$).
%
%
This assumption can be approximately satisfied when the input signal
is strongly chaotic and the sampling interval $T$ is sufficiently
large.

Because the probability distribution of sampled data is far from
uniform as shown in Fig. \ref{fig:histogram}(a), we expect the
probability of each $A_j$ (here we suppress the sample number index
$n$) taking the value $0$ (or $1$) to be biased.
We define the bias of the $j$-th bit as the deviation from 1/2 of the
probability that $A_j$ takes the value 0, i.e.,
\begin{equation}
\mbox{Bias}_j:=\left|\,\mbox{Prob}\left\{ A_j=0 \right\}-\frac{1}{2}\,
\right|.
\label{eq:bias}
\end{equation}
Figure \ref{fig:bias} (filled circles ($\bullet$)) shows the
$j$-dependence of the biases.
For this calculation, we used consecutive 1-Gbit samples.
In Fig. \ref{fig:bias}, we can see that a bit of less significance
has a smaller bias.
Such a tendency has been observed in previous studies \cite{Argyris10,
  Akizawa12, Oliver13}, although the reason for this tendency has not
been theoretically discussed.
In the Appendix, we provide a simple explanation for this tendency
based on dynamical systems theory.

Various bit manipulation methods for reducing the bias have been
examined in previous studies \cite{Reidler09, Hirano09, Argyris10,
  Akizawa12, Takahashi14, Sakuraba15}, which consist mainly of
exclusive-OR (XOR) operation, bit order reversing, and use of a
delayed signal.
XOR operation plays a key role in the bias reduction.
As is well known, when the random variables $X$ and $Y$ have biases
$\epsilon_X$ and $\epsilon_Y$, respectively, their XORed value $Z$
$:=$ $X$ $\bigoplus$ $Y$ has a reduced bias of
$2\epsilon_X\epsilon_Y$.

In the work we report in this paper, we employed bit manipulation
essentially similar to the ones used in \cite{Akizawa12, Takahashi14,
  Sakuraba15}.
In contrast to the previous studies, we perform the bit manipulation
in real time.
In our RBS system, the bit manipulation can be carried out either on
the FPGA of the ADC board or on a PC.
We adopted the latter, because it provides easier coding and more
tuning flexibility, although the former is more efficient in terms of
data transfer (namely, fewer data are transferred through the PCI
express bus when bit extraction is performed at the FPGA).

Recently, Uchida et al. \cite{Ugajin17} implemented a more complicated
bit manipulation algorithm in an FPGA and achieved real-time physical
RBG at up to 21.1 Gbps with ADC sampling rate of 3.6 GSps and bit
manipulation extracting 8-bits per sample.
However, in this case, the theoretical peak RBG rate exceeds the
actual RBG rate.
In the system reported in this paper, we prioritize the full description
of the end-to-end entropy source model describing how the entropy from
physical sources is converted into bit entropy, so we limit the
streaming rate to ensure continuous streaming.

\begin{figure}[!b]
\centering
\includegraphics[width=10cm]{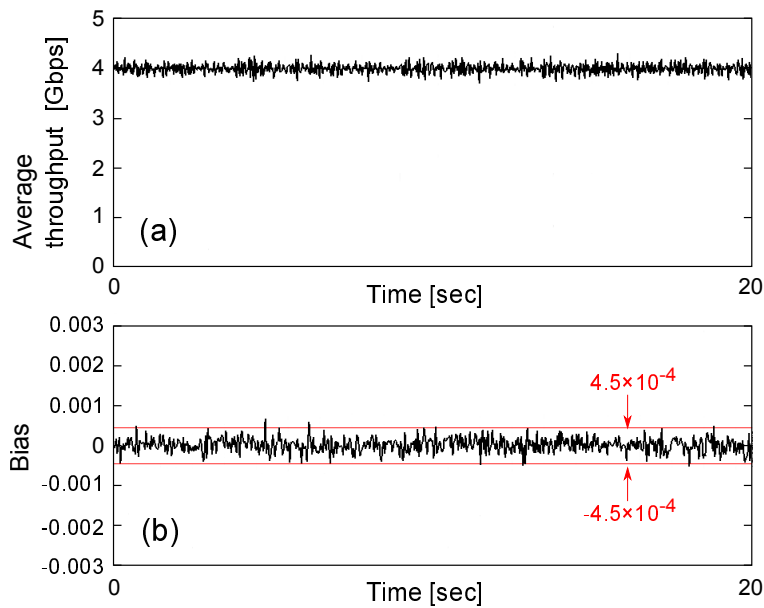}
\caption{Real-time monitoring of the average throughput and bias of
  the generated random bits, where the parameters for the bit
  manipulation were set at $m$ $=$ 4 (four-bit extraction) and $s=0$,
  yielding a generation rate of 4 Gbps.
  In (a), the throughput was measured in terms of the number of random
  bits supplied to the PC's user memory space per time.
  In (b), the bias was calculated for 8-Mbit samples.
  The two lines at $\pm$ $1.29/\sqrt{8\times 1024^2}$ $\approx$ $\pm
  4.5\times 10^{-4}$ correspond to the significance level $\alpha$ $=$
  0.01 for the null hypothesis that the generated bits are random.
}
\label{fig:monitoring}
\end{figure}

For the input for the bit manipulation, we use two 8-bit data sets,
$X(n T)$ and $X((n-n_d) T)$, where an integer $n_d$ is a delay
parameter that should be chosen so that the correlation between $X(n
T)$ and $X((n-n_d) T)$ is negligible.
We fixed $n_d$ $=$ 32, and the delayed data set was obtained by
software using the PC's memory.
We denote the binary expression of $X((n-n_d) T)$ as
$\{B_j\}_{j=0}^7$ (we suppress the $n$ dependence).
The output of the bit manipulation is an $m$-bit sequence ($m$ $\leq$
$8$) obtained as follows:
In the first step, we extract two sets of $m$-bit sequences,
$\{A_{s+k}\}_{k=0}^{m-1}$ and $\{B_{s+k}\}_{k=0}^{m-1}$, from
$\{A_j\}_{j=0}^7$ and $\{B_j\}_{j=0}^7$, where we introduced the
offset parameter $s$ $\in$ $[0,7]$ defining the lowest bit to be used
for RBG.
In the second step, we reverse the order of $\{B_{s+k}\}_{k=0}^{m-1}$
to obtain $\{B'_{s+k}\}_{k=0}^{m-1}$, where $B'_{s+k}=B_{s+m-1-k}$
($k$ $=$ 0, $\cdots$, $m-1$).
Finally, in the third stage, we perform a bitwise XOR operation for
$\{A_{s+k}\}_{k=0}^{m-1}$ and $\{B'_{s+k}\}_{k=0}^{m-1}$ to obtain
$C_{k}^{(s)}$ $=$ $A_{s+k}$
$\bigoplus$ $B'_{s+k}$ ($k$ $=$ 0, $\cdots$, $m-1$).

When we consider the empirical fact that a bit of less significance
has a smaller bias (c.f., Fig. \ref{fig:bias}), it is efficient to
perform the bit order reversal operation in the second step to obtain
as many less biased bits as possible, with minimal computational cost
to maintain the high throughput.
In Fig. \ref{fig:bias} (black squares ($\blacksquare$)), the effect of
the bit operation is illustrated for one-bit extraction (i.e., $m$ $=$
1) with $s$ $=$ 0, $\cdots$, 7, where we can confirm that the biases
of the lowest five bits can be made very small ($\lessapprox$
$10^{-5}$) by the XOR operation.

The bit manipulation algorithm was implemented in the random bit
server software on the PC (see Fig. \ref{fig:setup}(a)).
For an efficient implementation of the bit manipulation described
above, we used commands supported by SSE4.2 (Streaming SIMD Extensions
4.2), AMI (advanced bit manipulation), and BMI2 (Bit Manipulation
Instructions Sets 2).
By efficiently performing bit operations using these commands, we were
able to achieve a sustained streaming throughput up to 4 Gbps.

Figure \ref{fig:monitoring}(a) shows a real-time monitoring result for
the average throughput of random bits used by a user application,
where the average was obtained with 1/30th of a second time intervals.
We can confirm that the throughput seen by the user application is
sustained at the theoretical RBG rate of 4 Gbps.
This means that there is no bit loss during transfer to the PC's
shared memory space, and even to the user application.
Further, this means that there is no unknown factors in the bit
processing and transfer, allowing us to have a full description of the
process converting physical entropy to bit entropy.

Figure \ref{fig:monitoring}(b) shows the corresponding time series of
the bias, where 8-Mbit samples were used for the calculation of the
bias.
The two lines at $\pm$ $1.29/\sqrt{8\times 1024^2}$ $\approx$ $\pm
4.5\times 10^{-4}$ correspond to the significance level $\alpha$ $=$
0.01 for the null hypothesis that the generated bits are random.
For this demonstration, we used a bit manipulation algorithm with $m$
$=$ $4$ and $s$ $=$ $0$ (yielding a generation rate of 4 Gbps).
From this result, we can confirm a reasonable fluctuation of the bias
for the generated random bits.
In Sect. \ref{sect:NIST}, we present detailed information on the
quality of the generated random bits.

\section{Quality assessment of generated physical random bits}
\label{sect:NIST}
The statistical quality of physical random bits has been assessed by
using statistical test suites for random and pseudorandom number
generators, a representative of which is the NIST statistical test
suite known as NIST SP800-22 \cite{SP800-22}.
In previous studies, passing all the included tests has been adopted
as a criterion for generated bits to have acceptable statistical
quality.
Such a test suite has usually been applied to a limited number of
generated bit sequences.
For example, for NIST SP800-22, 1000 binary sequences with lengths of
$10^6$ bits (the total data amount is about 1 Gbits) are often used,
and the significance level $\alpha$ for $p$-values is set at 0.01,
following the recommendation in NIST SP800-22 \cite{SP800-22}.
Most previous studies provided information only for a single
trial of the NIST test suite.
However, considering the statistical nature of the NIST test suite and
the fact that the stationarity of the physical RBG system itself is
not trivial, the pass rate of the test suite would provide more
detailed information on the capability of the physical random bit
generator.
Such a pass rate has been evaluated for various representative
pseudorandom bit generators \cite{Okutomi10, Yamaguchi10, Lihua15}.
For example, \cite{Yamaguchi10} reported a pass rate of 41 \% to 56
\%, while \cite{Lihua15} reported a pass rate of 61 \% to 69 \%.
In the former, the NIST test suite was applied to 100 different sample
data sets to evaluate the pass rate.
By taking account of the correlations of the sub-tests included in the
NIST test suite, \cite{Lihua15} estimated the upper bound of the pass
rate to be 80.99 \%.

However, such a pass rate has not been systematically evaluated for
physical random bits.
In Sect. \ref{sect:pass_rate}, we evaluate the pass rate of the NIST
test suite by applying it to 100 different sample data sets obtained
with our RBS system, and reveal how this rate depends on the parameters
of the bit manipulation.
In addition, in Sect. \ref{sect:long_term}, we report the results of
long-term (60-minute) bias monitoring.

\begin{table}[!b]
\renewcommand{\arraystretch}{1.1}
\caption{The results of a single trial of the NIST test suite
  \cite{SP800-22} for a data set obtained when the bit manipulation
  parameters were set at $m$ $=$ 2 and $s$ $=$ 3. For 1000 binary
  sequences with lengths of $10^6$ bits and significance level
  $\alpha$ $=$ 0.01, each test is passed if the $P$-value (uniformity
  of $p$-values) is larger than 0.0001, and the proportion is in the
  0.99 $\pm$ 0.0094392 range. For the tests with multiple sub-tests,
  the worst $P$-values and proportions are shown.}
\centering
\footnotesize
\begin{tabular}{c|l|c|c|c}
\hline
No. & Test name (the number of sub-tests) & $P$-value & Proportion &
Result\\ \hline\hline
1 & Frequency (1) & 0.940080 & 0.9940 & Success\\ \hline
2 & Block Frequency (1) & 0.666245 & 0.9840 & Success\\ \hline
3 & Runs (1) & 0.605916 & 0.9840 & Success\\ \hline
4 & Longest Run (1) & 0.601766 & 0.9910 & Success\\ \hline
5 & Rank (1) & 0.034712 & 0.9840 & Success\\ \hline
6 & FFT (1) & 0.182550 & 0.9890 & Success\\ \hline
7 & Non-Overlapping Template (148) & 0.011144 & 0.9840 & Success\\ \hline
8 & Overlapping Template (1) & 0.699313 & 0.9920 & Success\\ \hline
9 & Universal (1) & 0.148653 & 0.9840 & Success\\ \hline
10 & Linear Complexity (1) & 0.707513 & 0.9950 & Success\\ \hline
11 & Serial (2) & 0.342451 & 0.9890 & Success\\ \hline
12 & Approximate Entropy (1) & 0.036833 & 0.9930 & Success\\ \hline
13 & Cumulative Sums (2) & 0.940080 & 0.9950 & Success\\ \hline
14 & Random Excursions (8) & 0.207821 & 0.9854 & Success\\ \hline
15 & Random Excursions Variant (18) & 0.003721 & 0.9968 & Success\\ \hline
\end{tabular}
\label{table:NIST_tests}
\end{table}

\subsection{Pass rate of NIST test suite}
\label{sect:pass_rate}
As described in Sect. \ref{sect:bit_manipulation}, the bit
manipulation algorithm has two parameters $m$ and $s$, where $m$ is
the number of extracted bits and $s$ the offset bit.
For various sets of $m$ and $s$, we applied the NIST test suite to 100
different sample data sets (each consisting of 1000 binary sequences
with lengths of $10^6$ bits), and evaluated the pass rate.
The NIST test suite consists of fifteen kinds of statistical tests,
and each tests a null hypothesis that the binary sequence is random
\cite{SP800-22}.
The fifteen kinds of the test and the number of sub-tests are listed
in the second column of Table \ref{table:NIST_tests}.
There are a total of 188 sub-tests.
The criterion for passing each sub-test is determined by the
significance level and the sequence length.
We set the significance level $\alpha$ for the $p$-values at $0.01$.
For 1000 binary sequences, the proportion of the binary sequence that
produces a $p$-value larger than $\alpha$ is expected to be in the
0.99 $\pm$ 0.0094392 range for ideal random bits.
For a given sub-test, if the proportion is in this expected range and
the $P$-value (uniformity of $p$-values) is larger than $0.0001$, we
conclude that the sample data set passes this sub-test, and for a
given test from the fifteen kinds listed in Table
\ref{table:NIST_tests}, if a sample data set passes all the included
sub-tests, we say that the sample data set passes the test.
Moreover, if a sample data set passes all the 188 sub-tests (or
equivalently passes all the fifteen kinds of the tests), we conclude
that it passes the NIST test suite.
In Table \ref{table:NIST_tests}, we show the results of a single trial
of the NIST test suite for a data set obtained when the bit
manipulation parameters were set at $m$ $=$ 2 and $s$ $=$ 3, where we
can confirm that all the fifteen tests were passed.

By using 100 sample data sets obtained by our RBS system, we evaluated
the pass rate for the NIST test suite as well as that for each of the
included fifteen kinds of the tests.
In Table \ref{table:NIST_results}, the counts of passing each of the
fifteen tests are summarized for various parameter values of the bit
manipulation algorithm, $m$ and $s$.
Since we have observed in Fig. \ref{fig:bias} that the bias is large
for $j$ $=$ 5, 6, and 7, we limited the values of $m$ and $s$, so that
these bits were not used for the RBG.
As given in the bottom row of Table \ref{table:NIST_results}, the pass
rates for the NIST test suite were in the 65 \% to 75 \% range, which
are comparable to the pass rates for commonly-used reliable
pseudorandom bit generators \cite{Yamaguchi10, Lihua15}.
Therefore, we conclude that the physical random bits generated by our
RBS system have sufficient statistical quality.

In Sect. \ref{sect:entropy_sources}, we found that the estimated
number of bits unaffected by the electronic noise is 5.42.
This means that the results for $s$ $=$ 0, 1, and 2 are suspected of
being more or less affected by the electronic noise.
In Table \ref{table:NIST_results}, we used superscript ($*$) for the
pass rates when the lowest three bits were not used in the RBG.
As far as the pass rate is concerned, we observed no significant change
caused by the inclusion of the electronic noise.
However, when simply characterizing the entropy production capability
of a chaotic laser, or for applications that demand high reliability
in an entropy source, one must exclude those bits that might be
contaminated by electronic noise whose characteristics are
unknown.

\begin{table}[!b]
\renewcommand{\arraystretch}{1.1}
\caption{The counts for passing each test in the NIST test suite for
  100 sample data sets, each consisting of 1000 binary sequences with
  lengths of $10^6$ bits. Each column corresponds to a different bit
  manipulation condition designated by $m$ and $s$, which respectively
  represent the number of extracted bits and the offset bit. The
  bottom row shows the pass rates for the NIST test suite. A pass rate
  with $*$ indicates that those bits suspected of being affected by
  electronic noise were not used in the RBG.}
\centering
\footnotesize
\begin{tabular}{c|c|c|c|c|c|c|c|c|c|c|c}
\hline
Test  & \multicolumn{5}{c|}{$m$ $=$ 1} & \multicolumn{4}{|c|}{$m$ $=$ 2} & \multicolumn{2}{|c}{$m$ $=$ 4}\\ \cline{2-12}
No. & $s$ $=$ 0 & $s$ $=$ 1 & $s$ $=$ 2 & $s$ $=$ 3 & $s$ $=$ 4 & $s$ $=$ 0 & $s$ $=$ 1 & $s$ $=$ 2 & $s$ $=$ 3 & $s$ $=$ 0 & $s$ $=$ 1\\ \hline
1 & 100 & 100 & 100 & 99 & 100 & 100 & 100 & 100 & 100 & 100 & 100\\ \hline
2 & 100 & 100 & 100 & 100 & 100 & 100 & 100 & 100 & 100 & 100 & 100\\ \hline
3 & 99 & 100 & 100 & 100 & 100 & 99 & 99 & 100 & 100 & 100 & 99\\ \hline
4 & 99 & 99 & 100 & 99 & 100 & 100 & 100 & 100 & 100 & 100 & 100\\ \hline
5 & 100 & 100 & 100 & 100 & 100 & 100 & 100 & 100 & 100 & 100 & 100\\ \hline
6 & 99 & 99 & 97 & 98 & 97 & 98 & 100 & 99 & 100 & 98 & 98\\ \hline
7 & 82 & 75 & 78 & 83 & 75 & 74 & 76 & 81 & 75 & 85 & 74\\ \hline
8 & 100 & 100 & 99 & 99 & 99 & 99 & 100 & 99 & 100 & 99 & 99\\ \hline
9 & 98 & 97 & 99 & 99 & 100 & 99 & 99 & 99 & 99 & 100 & 99\\ \hline
10 & 100 & 100 & 100 & 100 & 98 & 100 & 99 & 100 & 100 & 100 & 100\\ \hline
11 & 100 & 100 & 100 & 100 & 100 & 100 & 100 & 100 & 100 & 100 & 100\\ \hline
12 & 100 & 99 & 100 & 100 & 100 & 100 & 100 & 100 & 99 & 100 & 100\\ \hline
13 & 99 & 100 & 100 & 100 & 100 & 100 & 99 & 100 & 100 & 100 & 99\\ \hline
14 & 99 & 99 & 95 & 97 & 96 & 98 & 98 & 99 & 95 & 94 & 96\\ \hline
15 & 98 & 97 & 97 & 98 & 99 & 95 & 98 & 98 & 99 & 97 & 98\\ \hline\hline
Pass rate (\%) & 73 & 68 & 70 & 75$^*$ & 65$^*$ & 65 & 69 & 75$^*$ & 69$^*$ & 75 & 65\\ \hline
\end{tabular}
\label{table:NIST_results}
\end{table}

\subsection{Long-term bias monitoring}
\label{sect:long_term}
To confirm the operational stability of our RBS system, we monitored
the bias of generated random bits for a long time interval of 60
minutes.
For the bit manipulation, we set $m$ $=$ 1 (one-bit extraction) and
$s$ $=3$, which yields a 1 Gbps generation rate.
The bias monitor computed the bias in 8 Mbit blocks acquired from the
PC's shared memory space.

\begin{figure}[!t]
\centering
\includegraphics[width=14cm]{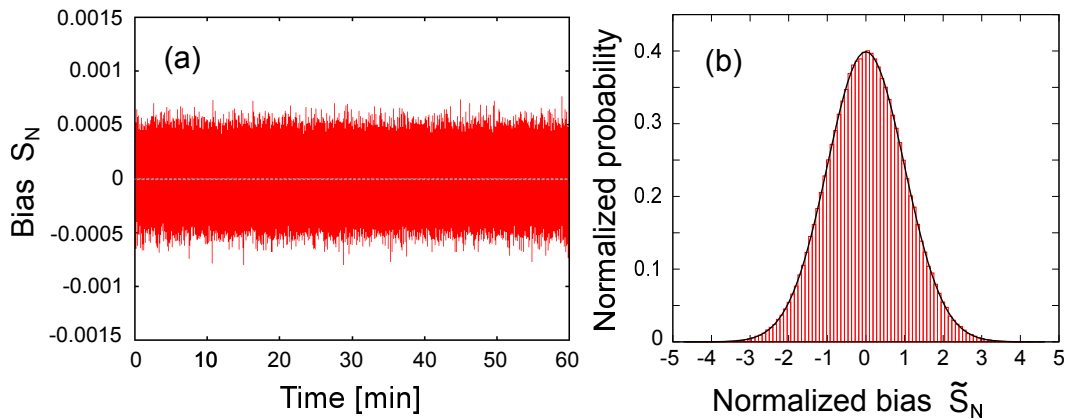}
\caption{(a) Time dependence of the bias $S_N$ for generated random
  bits. The bias $S_N$ is calculated from $N$ ($=$ $8$ $\times$
  $1024^2$) consecutive bits. (b) Normalized probability distribution
  of the scaled bias $\tilde{S}_N$ for the data in (a) (see text for
  definitions of $S_N$ and $\tilde{S}_N$), where the solid curve is
  the normalized Gaussian distribution (Eq. (\ref{eq:gaussian}))
  predicted by the central limit theorem.}
\label{fig:bias_monitoring}
\end{figure}

For a generated bit sequence of length $N$ $=$ $8\times 1024^2$,
$\{z_j\}_{j=1}^N$, we calculated the following (signed) bias:
\begin{equation}
S_N=\frac{1}{N}\sum_{k=1}^N z_k-\frac{1}{2}.
\end{equation}
The calculation was carried out in real time, and repeated with an
average time interval of 8.39 $\mu$s.
This time interval is slightly longer than that needed for generating
an $N$ ($=$ 8 M) bit sequence with a 1 Gbps rate, estimated as 7.81
$\mu$s.
During the 60 minutes, we obtained about 4.29 $\times$ $10^5$ samples
for $S_N$.
Figures \ref{fig:bias_monitoring}(a) and \ref{fig:bias_monitoring}(b)
respectively show the time dependence of $S_N$ and the normalized
probability distribution $P(\tilde{S}_N)$ for the scaled bias
$\tilde{S}_N$ $=$ $2\sqrt{N}S_N$.
From Fig. \ref{fig:bias_monitoring}(a), we can confirm the stable
operation of our RBS system.
According to the central limit theorem, $\tilde{S}_N$ is expected to
obey a normalized Gaussian distribution
\begin{equation}
P(\tilde{S}_N)=\frac{1}{\sqrt{2\pi}}e^{-\tilde{S}_N^2/2}.
\label{eq:gaussian}
\end{equation}
In Fig. \ref{fig:bias_monitoring}(b), this theoretical prediction is
shown by a bold curve, and it provides an excellent fit with the
actual data obtained with the long-term bias monitor.
This result further convinces us the reliability of our RBS system.

\section{Conclusion}
We demonstrated a physical random bit streaming (RBS) system with a
chaotic laser as its physical entropy source.
The RBS system is designed to allow a full description of how the
entropy from physical sources is converted to bit entropy.
The system can both perform bit manipulation for bias reduction in
real time and continuously supply random bits to the shared memory
space of the PC.
We confirmed a throughput of up to 4 Gbps for the supply of the
physical random bits.
These physical random bits are ready to use in the sense that their
bias is made negligible.
The statistical quality of the generated random bits was
systematically assessed by using the NIST test suite.
Using a large number of generated physical random bits, we evaluated
the pass rate of the NIST test suite to be 65 \% to 75 \%, which is
comparable to the pass rates previously evaluated for commonly-used
reliable peudorandom bit generators \cite{Okutomi10, Yamaguchi10,
  Lihua15}.
We also confirmed the long-term operational stability of our RBS
system, by monitoring the bias for 60 minutes.
This long-term continuous monitoring result proved that the chaotic
laser chip can stably generate sufficiently random signals, even under
unavoidable temperature and current fluctuations.
This point has not been verified in previous studies
\cite{Sunada12,Takahashi14,Harayama11}, but is important for
confirming the reliability of the chaotic laser chip as an entropy
source.

\section*{Appendix: Bias dependence on bit number}
\label{sect:bernoulli}
\begin{figure}[!b]
\centering \includegraphics[width=13cm]{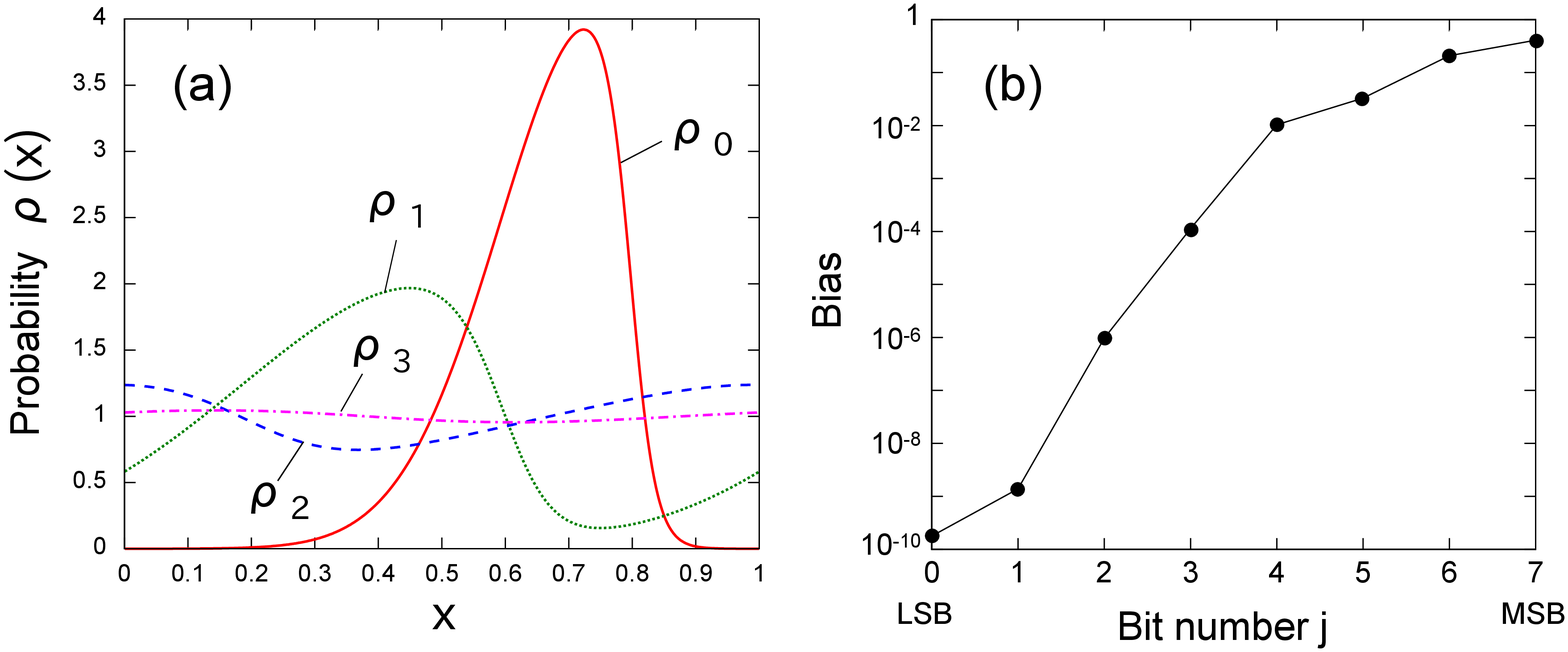}
\caption{(a) Evolution of $\rho_0(x)$ under the Bernoulli map,
  Eq. (\ref{eq:bernoulli}). As $j$ increases, $\rho_j(x)$ converges to
  a uniform distribution $\rho_{\infty}(x)$ $\equiv$ $1$.
(b) Bias dependence on bit number $j$, where $j$ $=$ $0$ corresponds
  to the LSB.
}
\label{fig:bernoulli}
\end{figure}
In Fig. \ref{fig:bias} (filled circles ($\bullet$)), we can observe a
tendency for a bit of less significance to have a smaller bias.
In this Appendix, we explain the reason for this tendency.
Let us use $X_0$ to denote a (decimal) variable describing sampled
data.
For simplicity and without loss of generality, we assume that $X_0$
$\in$ $[0,1)$, and that $X_0$ obeys the probability distribution
  $\rho_0(x)$.
The decimal value of $X_0$ can be expressed by the binary form $.A_0
A_1 A_2 \cdots$, where $A_j\in\{0,1\}$ $(j=0,1,2,\cdots)$ are
determined so as to satisfy $X_0=\sum_{j=0}^{\infty} A_j 2^{-(j+1)}$.
In dynamical systems theory \cite{Beck-Schloegl}, it is well known
that the binary shift operation $.A_0 A_1 A_2 \cdots$ $\mapsto$ $.A_1
A_2 A_3 \cdots$ is equivalent to the mapping of the point $X_0$ by the
Bernoulli map $B: [0,1) \mapsto [0,1)$ defined by
\begin{equation}
B(x)=2x~\mbox{mod}~1=
\left\{
\begin{array}{ll}
2x & \mbox{for}~x\in [0,\frac{1}{2})\\
2x-1 & \mbox{for}~x\in [\frac{1}{2},1),
\end{array}
\right.
\label{eq:bernoulli}
\end{equation}
where the mapped point $X_1=B(X_0)$ has the binary expression $.A_1
A_2 A_3 \cdots$.
Using this fact, we can write the probability of the $j$-th bit $A_j$
taking the value 0 as
\begin{equation}
\begin{array}{ll}
\mbox{Prob}\{A_j=0\} &= \mbox{Prob}\{B^j(x_0)< 1/2\}\\
&= {\displaystyle \int_{0}^{1/2} dx \rho_j(x)},~\quad(j=0,1,\cdots)
\end{array}
\label{eq:prob}
\end{equation}
where $\rho_j(x)$ is the evolution of $\rho_0(x)$ obtained by the
$j$-time iteration of the Bernoulli map, and it can be written as
$\rho_j(x)={\cal F} \rho_{j-1}(x)$ by using the Frobenius-Perron
operator ${\cal F}$ defined by
\begin{equation}
{\displaystyle
{\cal F}\rho_{j-1}(x):=\frac{1}{2}\left[
\rho_{j-1}\left(\frac{x}{2}\right)+\rho_{j-1}\left(\frac{x+1}{2}\right)
\right]
}.
\label{eq:FP_operator}
\end{equation}
From Eq. (\ref{eq:FP_operator}), we have
\begin{equation}
\rho_j(x)=\frac{1}{2^j}\sum_{m=0}^{2^j-1}\,\rho_0\left(\frac{x+m}{2^j}\right).
\label{eq:rho_j}
\end{equation}
Equation (\ref{eq:rho_j}) with $j$ $\to$ $\infty$ is nothing but a
Riemann integral of $\rho_0(x)$, and the R.H.S. of
Eq. (\ref{eq:rho_j}) converges to $\int_{0}^{1}dx \rho_0(x)$ $=$ $1$
for $j$ $\to$ $\infty$ when $\rho_0(x)$ is integrable \cite{Driebe}.
In other words, $\rho_j(x)$ approaches a uniform distribution
(natural invariant measure) $\rho_{\infty}(x)\equiv 1$ as $j$ goes to
infinity.
From Eq. (\ref{eq:prob}), it directly follows that
$\mbox{Prob}\{A_j=0\}$ goes to $1/2$ as $j$ goes to infinity.
Thus, the bias of the $j$-th bit defined in Eq. (\ref{eq:bias}) goes
to zero as $j$ goes to infinity.

In Fig. \ref{fig:bernoulli}(a), we numerically illustrate the
convergence of $\rho_j(x)$ to a uniform distribution, when the
initial distribution $\rho_0(x)$ is given by a skewed unimodal
distribution.
In Fig. \ref{fig:bernoulli}(a), we observe that $\rho_j(x)$
approaches $\rho_{\infty}(x)\equiv 1$ as $j$ increases.
In Fig. \ref{fig:bernoulli}(b), we plot the bias dependence on the bit
number $j$, where the biases were calculated from $\rho_j(x)$ using
Eqs. (\ref{eq:bias}) and (\ref{eq:prob}).
The biases exhibit exponential decay as a function of $j$, which is
due to the smoothness of the probability distribution $\rho_0(x)$.
For real experimental data, the probability distribution $\rho_0(x)$
often contains an ever finer fluctuation that remains at the level of
the LSB, which may lead to non-exponential decay or decay saturation.

\begin{acknowledgments}
We thank Tomohiro Miyasaka and Fumihiro Matsui for coding the programs
for the bit manipulation and random bit streaming.
S.S and K.A. thank Atsushi Uchida for sending them a preprint on an
FPGA implementation of real-time physical RBG \cite{Ugajin17}, Dai
Ikarashi, Koji Chida and Hitoshi Ito for discussions regarding secret
sharing, Hiroyuki Noto for discussions concerning signal processing,
and Tomohiro Nakatani for his support and encouragement.
\end{acknowledgments}

\end{document}